\documentclass[seceq,supplement]{ptptex}

\usepackage{graphicx}
\usepackage{wrapft}

\newcommand{\bfk}{\mbox{{\boldmath $k$}}}

\newcommand{\beq}{\begin{eqnarray}}
\newcommand{\eeq}{\end{eqnarray}}
\newcommand{\btem}{\bibitem}

\newcommand{\la}{\langle}
\newcommand{\ra}{\rangle}

\newcommand{\dirac}{\partial\llap{$\diagup$\kern-2pt}}

\title{
QCD Critical Points and Their Associated Soft Modes%
}

\author{
T. \textsc{Kunihiro}$^1$,
Y.  \textsc{Minami}$^1$
and 
Z.  \textsc{Zhang}$^2$%
}

\inst{
$^1$Department of Physics, Kyoto University,
Sakyo-ku,  606-8502, Kyoto, Japan\\
$^2$
Mathematical \& Physical Science School, North China Electric Power  University, 
Zhuxinzhuang, Changping District,Beijin 102206, China
}

\abst{
The mean-field level calculation shows that the QCD matter can have  multiple
 critical points incorporating the color superconductivity under charge neutrality
constraint due to  the repulsive vector interaction; this actually
implies  that the QCD matter is very soft for a simultaneous formation
of  diquark and chiral condensates coupled with the baryonic density.
Dynamical density fluctuations are analyzed as possible soft modes
around the QCD critical point using dissipative relativistic fluid dynamics.
It is found that  the entropy fluctuation solely  gets enhanced while
the sound modes due to mechanical density fluctuations are 
strongly attenuated around the QCD CP, 
which  may suggest a suppression or even total disappearance 
of  Mach cone at the CP.
}

\begin{document}

\maketitle

\section{Introduction}

A unique feature of the phase diagram of quantum chromodynamics (QCD)
 is the existence of a critical point (CP)\cite{Asakawa:1989bq,Barducci:1989};
see, for a review,\citen{Stephanov:2007fk}.
In the first part of this talk, however, we shall describe
possible variants of the QCD phase diagram  when the color superconductivity(CSC) is
taken into account, focusing on a  possible important role of 
the vector-vector interaction between quarks\cite{Kitazawa:2002bc,Fukushima:2008is,Zhang:2009mk}.
The mean-field level calculation shows that there can exist multiple critical points
in the QCD phase diagram when the repulsive vector
interaction is taken into account \cite{Kitazawa:2002bc,Zhang:2009mk}, which feature
can be enhanced when the charge neutrality constraint is imposed\cite{Zhang:2008wx};
see also \citen{Hatsuda:2006ps} for a similar result in
a different context.
The possible emergence of multiple critical points  actually implies that 
the QCD matter around the phase boundary of the
chiral-to-CSC transition is soft for a combined excitation of the baryon density, scalar condensate
and the diquark excitations.

Around a critical point of a second-order transition,
  large fluctuations of physical quantities are expected.
It has been established \cite{fujii,son} that 
the QCD CP belongs to the same universality class($Z_2$) as the liquid-gas CP, 
 and  hydrodynamic modes coupled to conserved quantities such 
as sound modes are  softening modes at the CP:
We show on the basis of the dissipative relativistic
fluid dynamics 
that the dynamical density fluctuations or the Brillouin modes are greatly suppressed
around the CP, while the energy fluctuations or the Rayleigh modes
 are enhanced when approaching the CP\cite{Minami:2009hn}.

We shall conclude the report by  mentioning
 some other examples of phase transitions that the QCD matter may undergoes
and the respective soft modes, which could affect the quark spectrum drastically.

\section{QCD critical points; 
alternatives of QCD phase diagram with vector interaction and charge neutrality
 }

When charge neutrality is imposed to the quark matter,
a mismatch of the Fermi surfaces of the respective flavors arises unless
the chiral limit is taken\cite{Alford:2002kj}.
 For  such a system, it is known that the diquark gap has an abnormal 
temperature dependence that the gap 
increases as the temperature is raised, and has the maximum 
at a finite temperature\cite{Shovkovy:2003uu}:
This  is due to the smearing of the Fermi surface by
temperature. On the other hand,
 such a system 
tends to become unstable due to the color magnetic instability\cite{Huang:2004bg}.

It is known that the repulsive vector 
interaction~\cite{ref:ES,Asakawa:1989bq,ref:KLW,ref:Kuni2,ref:Bub,ref:BHO}
 in  the Nambu-Jona-Lasinio(NJL) model\cite{NJL}
as a low-energy effective model of QCD \cite{ref:Kle,ref:HK,BuballaReview} 
postpones the chiral restoration towards larger
chemical potential\cite{Asakawa:1989bq,ref:KLW,ref:BHO}.
We note that the renormalization-group analysis
~\cite{ref:EHS,ref:SW-reno}, the chiral instanton-anti-instanton
molecule model ~\cite{ref:SS} and the truncated Dyson-Schwinger
model of QCD ~\cite{ref:RWP} all support the existence of the
vector-vector four-quark interaction.
When the possible transition to a CSC phase is
considered in the NJL model, it was found that there can appear another critical point 
in the QCD phase diagram~\cite{Kitazawa:2002bc}.
The relevance of the repulsive vector-vector interaction
to the chiral and CSC transitions
 can be intuitively understood as follows\cite{ref:kuni_conf}.
According to thermodynamics, when  two phases I and II 
are in an equilibrium state, 
their temperatures $T_{\rm I, II}$, pressures $P_{\rm I, II}$ and 
the chemical potentials $\mu_{\rm I, II}$ are the same:
$T_{\rm I}=T_{\rm II}$,\,
$P_{\rm I}=P_{\rm II}$,\, and 
$\mu_{\rm I}=\mu_{\rm II}$.
If the two phases are the chirally broken and  restored phase 
with quark masses satisfying $M_I>M_{II}$, 
the last equality implies that
the chirally restored phase has a higher density than the
broken phase, because $\mu_{\rm I, II}$ 
at vanishing temperature are expressed as $\mu_{\rm i}=\sqrt{M_i^2+p_{F_i}^2}$, ($i=$ I, II),
and hence $p_{F_I}<p_{F_{II}}$, 
where  $p_{F_i}$ is the Fermi momentum of the $i$-th phase.
This means  that chiral restoration at finite density is necessarily
accompanied by a density jump to a higher density state with a
large Fermi surface, which in turn favors the formation of Cooper
instability leading to CSC.
However, since the vector coupling in the zero-th component
couples to the quark (or baryon) density $\rho_B=\la \bar{q}\gamma^0q\ra$,
it gives rise to a repulsive energy proportional to the density squared, i.e.
$G_{V}\rho_B^2/2$ with $G_V$ being the vector coupling.
This means that the restored phase is disfavored energetically due to the
vector coupling, and hence the vector coupling weakens and delays the phase 
transition of the chiral restoration at low temperatures.
Thus one also expects that the vector interaction
postpones  the formation of CSC 
to higher chemical potentials.

Zhang, Fukushima and Kunihiro\cite{Zhang:2008wx}
 found that the positive electric chemical potential $\mu_e$ inherent in the
charge neutrality constraint plays a similar role with the repulsive
vector interaction on the chiral phase transition in a four-quark
interaction model: In a simple two-flavor NJL model, the
same two-critical-point structure as found in
\citen{Kitazawa:2002bc} emerges. In addition, positive $\mu_e$
also represents  the magnitude of difference between the Fermi spheres
of u and d quarks when taking into account the local charge
neutrality constraint. For an asymmetric homogeneous system
with the mismatched Fermi spheres,
 the energy gap of the Cooper
paring between these two flavor quarks can {\em increase} 
with temperature\cite{Shovkovy:2003uu}, as mentioned before.
 For two-flavor neutral CSC phase, this unconventional
thermal behavior of the diquark condensate can lead to a 
competition between chiral condensate and diquark condensate,
which can be enhanced with increasing temperature. For some model
parameters region, this abnormal competition induced by $\mu_e$
can result in a phase structure with even  {\em three} critical points~\cite{Zhang:2008wx}.
Thus one would expect that a simultaneous incorporation of 
the repulsive vector interaction and the electric chemical potential
under neutrality constraint
will weaken more significantly the phase transition from the chiral-broken to CSC phase.

Recently, Zhang and Kunihiro\cite{Zhang:2009mk} explored the effect of
the repulsive vector-vector interaction combined with
electric-charge neutrality in $\beta$-equilibrium  on the chiral
phase and CSC phase transitions within both two-flavor and
two-plus-one-flavor NJL models.
For the two-flavor case with u and d quarks, a nonlocal NJL model
\cite{Alford:1997zt,Grigorian:2006qe} is adopted in \citen{Zhang:2009mk}:
The vector-vector term that is U(2)$\times$U(2) invariant is chosen as 
\beq
{\cal L}_V=-G_V \sum_{i=0}^3\left[ \left(
\bar{q}(x) \gamma^\mu \tau_i q(x) \right)^2 - \left( \bar q(x) 
\gamma^\mu \gamma_5 \tau_i q(x) \right)^2 \right],
\label{Lagrangian2f}
\end{eqnarray}
where
$q(x)=\int{dy^4\tilde{f}(x-y)\psi(y)}$.
 For other notations, 
we refer to \citen{Zhang:2009mk}.
We remark  that 
our choice of  the chiral-invariant vector part involves also the
axial vector part, and
the vector (and axial vector) terms $\sim (\bar{u}\gamma^{\mu}u)^2\,+\,(\bar{d}\gamma^{\mu}d)^2$
 have no flavor mixing term.
We should mention that the other chiral
invariant vector interaction $(\bar{q}\gamma^{\mu}q)^2$ as adopted in 
\cite{Kitazawa:2002bc} has 
a flavor mixing and may cause
 different density dependence of the phase diagram from those
given in the present work.
The nonlocal interaction is adopted to deal with the one loop
ultraviolet divergence for the calculation of the
Meissner mass squared at finite temperature\cite{Kiriyama:2006jp}. We 
take a Lorentzian-type form factor\cite{Grigorian:2006qe},\,
$f^2(p=|{\mathbf{p}}|)=g(p)=\frac{1}{1 + (\frac{p}{\Lambda})^{2a}}$,
with $a=10$,
where $f(p)$ is the Fourier transformation of the form factor
$\tilde{f}(x)$.  Oher three model parameters are determined by the vacuum physical
quantities of the pion mass $M_\pi=135\text{MeV}$, pion decay constant
$f_\pi=92.4 \text{MeV}$ and the quark condensate
$-\langle \bar{u}u\rangle ^{1/3}\approx {250} \text{MeV}$.

 We notice that the net quark chemical potential becomes dynamical as given by
${\tilde{\mu}}_\alpha(p) = \mu_\alpha-4G_V{\rho_\alpha} g(p)$,
where $\rho_{\alpha}$ being the quark density of the flavor $\alpha$.
It is to be noted that the {\em induced  chemical potential},
$-4G_V{\rho_\alpha} g(p)$ and hence the dynamical quark chemical potential
${\tilde{\mu}}_\alpha(p)$ for u
 and d quarks can be different  from each other because
the electric-charge neutrality makes $\mu_d$ larger than $\mu_u$ and hence $\rho_d>\rho_u$.
Thus the  mismatch between the effective chemical potentials of u
quark and d quark becomes
$\delta\tilde{\mu}=\tfrac{1}{2}(\mu_e-4G_V(\rho_d-\rho_u)g(p))$,
which tells us that 
the difference in the u and d quark densities in turn makes smaller 
the mismatch of their effective chemical potentials, thanks to the vector interaction.
 This effect is found to play an
important role for the stability against the color magnetic
instability\cite{Zhang:2009mk}.

\begin{figure}[hbtp]
\begin{tabular}{cc}
\begin{minipage}{0.5\hsize}
  \begin{center}
\includegraphics[width=7cm]{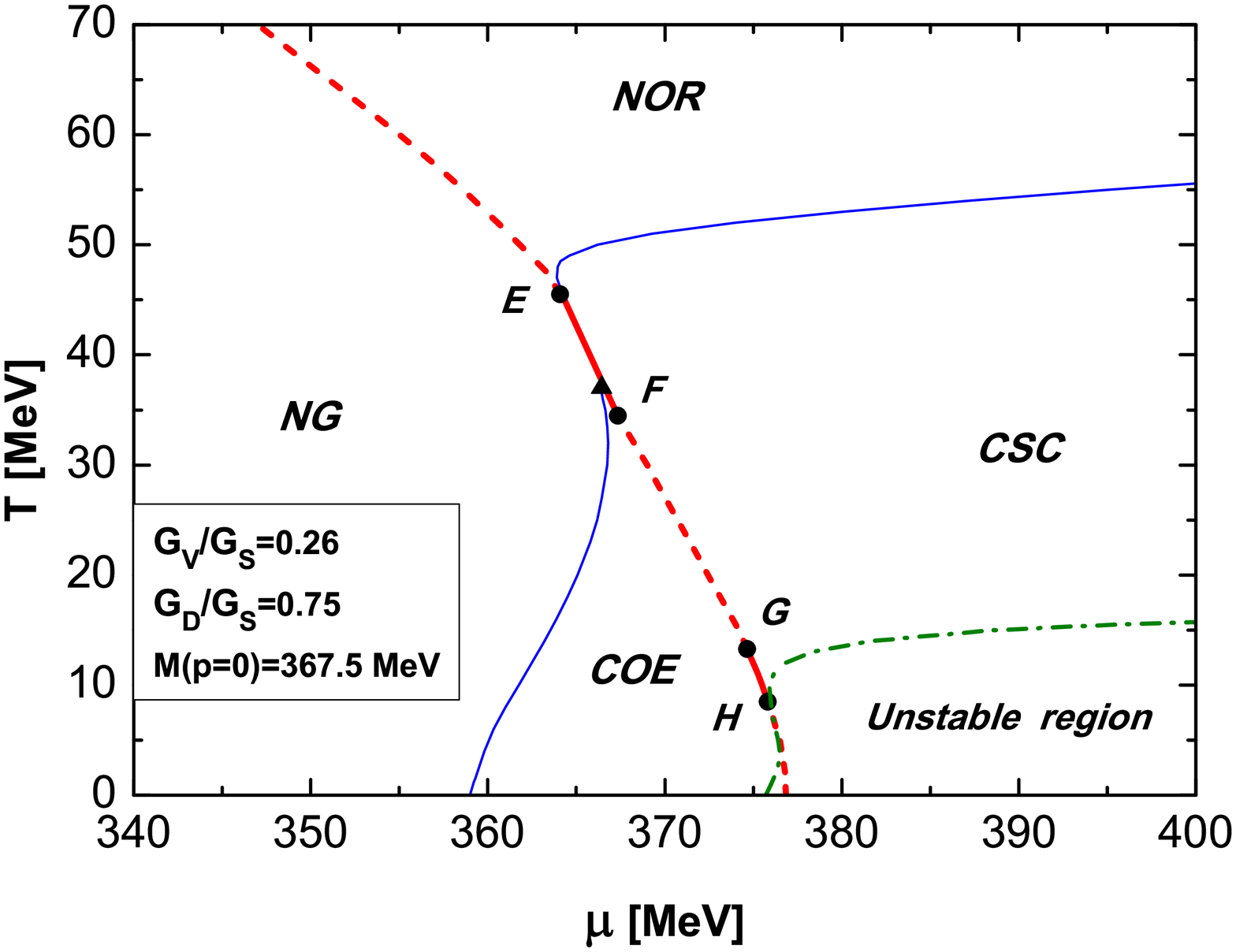}
\end{center}
\end{minipage}
\begin{minipage}{0.5\hsize}
 \begin{center}
\includegraphics[width=7cm]{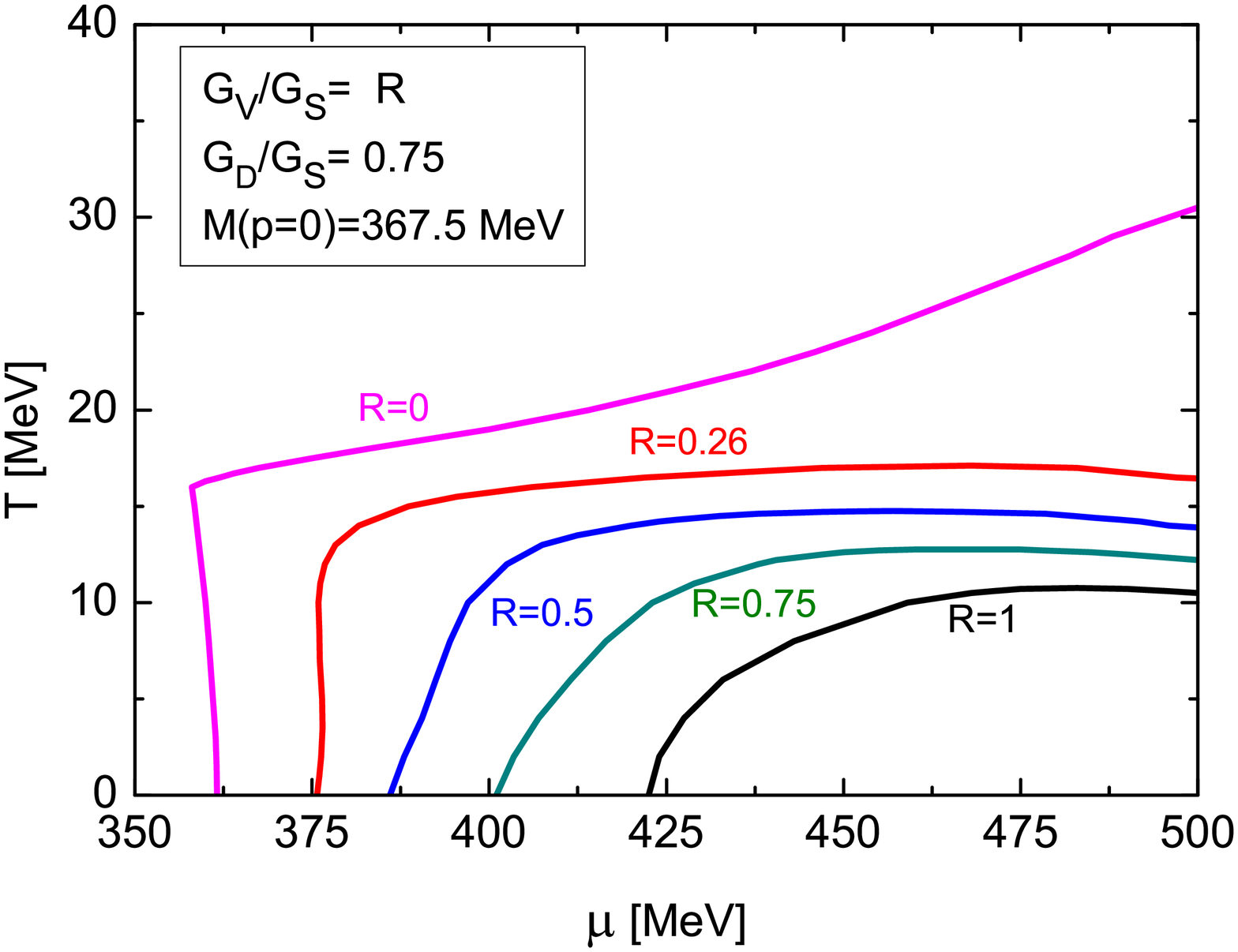}
\end{center}
\end{minipage}
\end{tabular}
\caption{
{\bf Left panel:}\, The phase diagram for  a model-parameter set with 
$G_V/G_S=0.26$ and fixed $G_D/G_S=0.75$ .  Here $G_S$ \, ($G_D$)
 refers to the scalar (diquark) coupling;
$G_S$ and other parameters are set to gives $M(p=0)=367.5$\, MeV. 
The unstable region with
chromomagnetic instability is indicated by the dash-dotted curve.
There appear four critical points being denoted by E,\,F,\, G\, and H.\,
{\bf Right panel:}\, The $G_v$ dependence of the 
the unstable region, which tends to shrink and move to the high density region
as $G_V$ is increased. Both taken from \citen{Zhang:2009mk}.} 
\label{fig:pdset2}
\end{figure}

In the left panel of Fig.~\ref{fig:pdset2}, we show an example of the phase diagram with
a finite vector coupling under the charge neutrality:
 We use the  abbreviations NG, CSC, COE,
and NOR refer to the hadronic (Nambu-Goldstone) phase with
$\langle \bar{q}_{\alpha}^aq_\alpha^a \rangle \equiv \sigma_{\alpha}\neq0$ and 
$\langle(\bar{q}_C)_{\alpha}^{a} i \gamma_5 \epsilon^{\alpha \beta3}\epsilon_{a b 3} q_{\beta}^{b} \rangle$
$\equiv \Delta=0$, the CSC phase with
$\Delta\neq0$ and $\sigma_{u,d}=0$, the coexisting phase with
$\sigma_{u,d}\neq0$ and $\Delta\neq0$, and the normal phase with
$\sigma_{u,d}=\Delta=0$, respectively, although they have exact meanings
only in the chiral limit.
We note that there appear four critical points, which are denoted 
by E, F, G and H.
With an increase of the vector coupling,
the lower two critical points, G and H disappear, while
the upper two critical points, F and G, remain  in the phase diagram.
 One should mention that 
 the critical point  H is located on the border between the stable
region and the unstable region, while other critical points are
free from the chromomagnetic instability.


A detailed analysis \cite{Zhang:2009mk} shows that
for the parameter set which gives $M(p=0)=367.5$\, MeV
 and the standard diquark coupling, five different types of
chiral critical point structures may exist,  and 
the number of the critical points changes as 
$1\,\rightarrow\,2\,\rightarrow\,4\,\rightarrow\,2\,\rightarrow\,0$ with 
the vector coupling being increased.

We should notice here that the
repulsive vector interaction also suppresses the magnitude of the
diquark condensate due to the reduced effective quark chemical
potential. However, the direct effect of the vector interaction on
$\delta\mu$ is more significant than that on $\Delta$, in
particular for finite temperature. Thus, as shown in 
the right panel of Fig.\ref{fig:pdset2}, we reach a significant 
findings that an increased vector coupling 
tends to shrink the unsable region toward the hifh-density and lower-temperature 
region. Needless to say, however,
 the vector interaction may not totally remove the unstable
region from the phase diagram
and hence other mechanism will  be still
necessary for a thorough cure of the magnetic instability.

\begin{figure}[htb]
\centerline{\includegraphics[width=7cm]{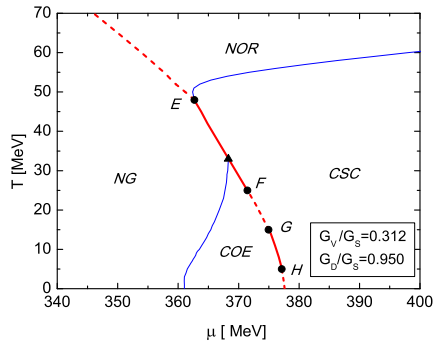}}
\caption{
A phase diagrams with four critical points for the
two-plus-one-flavor NJL model with  $G_V/G_S=0.312$ and 
$G_D/G_S$, under electric-charge-neutrality. Taken from \citen{Zhang:2009mk}.} 
\label{fig:two-plus-one}
\end{figure}

The above results obtained for the two-flavor
case is not essentially altered  even when the strange 
quark is taken into account\cite{Zhang:2009mk}, as shown in Fig.\ref{fig:two-plus-one}, where the  
 so-called Kobayashi-Maskawa-'tHooft(KMT) term\cite{KMT,ref:HK} is
also incorporated
\begin{eqnarray}
{\cal L}_{KMT}=
-K \left\{
\det_{f}\left[ \bar \psi \left( 1 + \gamma_5 \right) \psi \right]
+ \det_{f}\left[ \bar \psi \left( 1 - \gamma_5 \right) \psi
\right] \right\}.
 \label{Lagrangian2}
\end{eqnarray}

We should, however, remark 
that the number of the critical points is sensitive to the ratio 
$G_V/G_S$ and other parameters; other choice of the ratio and other parameters
would lead to different
phase structures with a varying number of the critical points\cite{Zhang:2009mk}.  
The message of such results we should have is that 
the QCD matter around the phase boundary is very soft for the density fluctuations combined with
 the formation of the
chiral and diquark condensates along the critical line 
when the color superconductivity is incorporated.

\section{Density and energy fluctuations around QCD CP}

Now apart from possible variants of the QCD phase diagram,
let us examine what would be good signatures or observables
that reflect the existence of the QCD CP.
Motivated by the fact that QCD CP belongs to
the same universality class as the liquid-gas transition,
Minami and Kunihiro \cite{Minami:2009hn}
 have analyzed the dynamical density fluctuations 
using various dissipative relativistic fluid dynamic equations irrespective of the
first or second order ones.
The dynamical structure factor (or spectral function) for the density fluctuation 
is calculated to be\cite{Minami:2009hn}
\begin{eqnarray}
\frac{S_{n n}(\bfk ,\omega )}{\la (\delta n(\bfk ,t=0))^2\ra}    = \frac{\gamma -1}{\gamma}
   \frac{2\Gamma_{\rm R} k^{2}}{\omega^{2}+\Gamma_{\rm R}^{2}k^{4}}
   +\frac{1}{\gamma}
   \{\frac{\Gamma_{\rm B} k^{2}}{(\omega -c_{s}k)^{2}+\Gamma_{\rm B}^{2}k^{4}}
+\, (\omega \rightarrow -\omega)\}
   \label{eq:landau}
\end{eqnarray}
where
$\Gamma_{\rm R}=\kappa/(n_0c_p)$ and
$\Gamma_{\rm B}=\frac{1}{2}[\Gamma_{\rm R}(\gamma-1)+\nu_l ]$
$+\frac{1}{2}c_s^2 T_0 (\frac{\kappa}{w_0} -2\Gamma_{\rm R} \alpha_P )$
with $\nu_l=(\zeta+4\eta/3)/w_0$ being the relativistic
longitudinal kinetic viscosity.
The enthalpy density in the equilibrium is denoted by $w_0$, while
$\gamma=c_p/c_v$ denotes the ratio of the specific heats at constant pressure and volume.
We refer to \citen{Minami:2009hn} for other notations.
 The spectral function has three peaks at frequencies $\omega=0$ and 
$\omega = \pm c_s k$, which
 corresponds to the entropy fluctuations (Rayleight peak),
 and mechanically induced density 
fluctuation(Brillouin peaks), respectively.
Notice that the pre-factor for the spectral function for the sound modes is
proportional to $1/\gamma$, which tends to vanish when approaching the CP;
$c_p$ behaves like $\xi^{2-\eta}$ in terms of the correlation length
$\xi$ and the critical exponent $\eta\sim 0.03$. 
Thus one sees that 
the mechanical density fluctuations are attenuated owing to the divergence of
the correlation length $\xi$ around the QCD CP.
On the other hand, the entropy fluctuation in turn is enhanced
and tends to makes a single peak around the QCD CP in the dynamical structure
factor of the density fluctuations.

\begin{figure}[htbp]
\begin{tabular}{cc}
\begin{minipage}{0.5\hsize}
  \begin{center}
   \includegraphics[width=40mm,angle=-90]{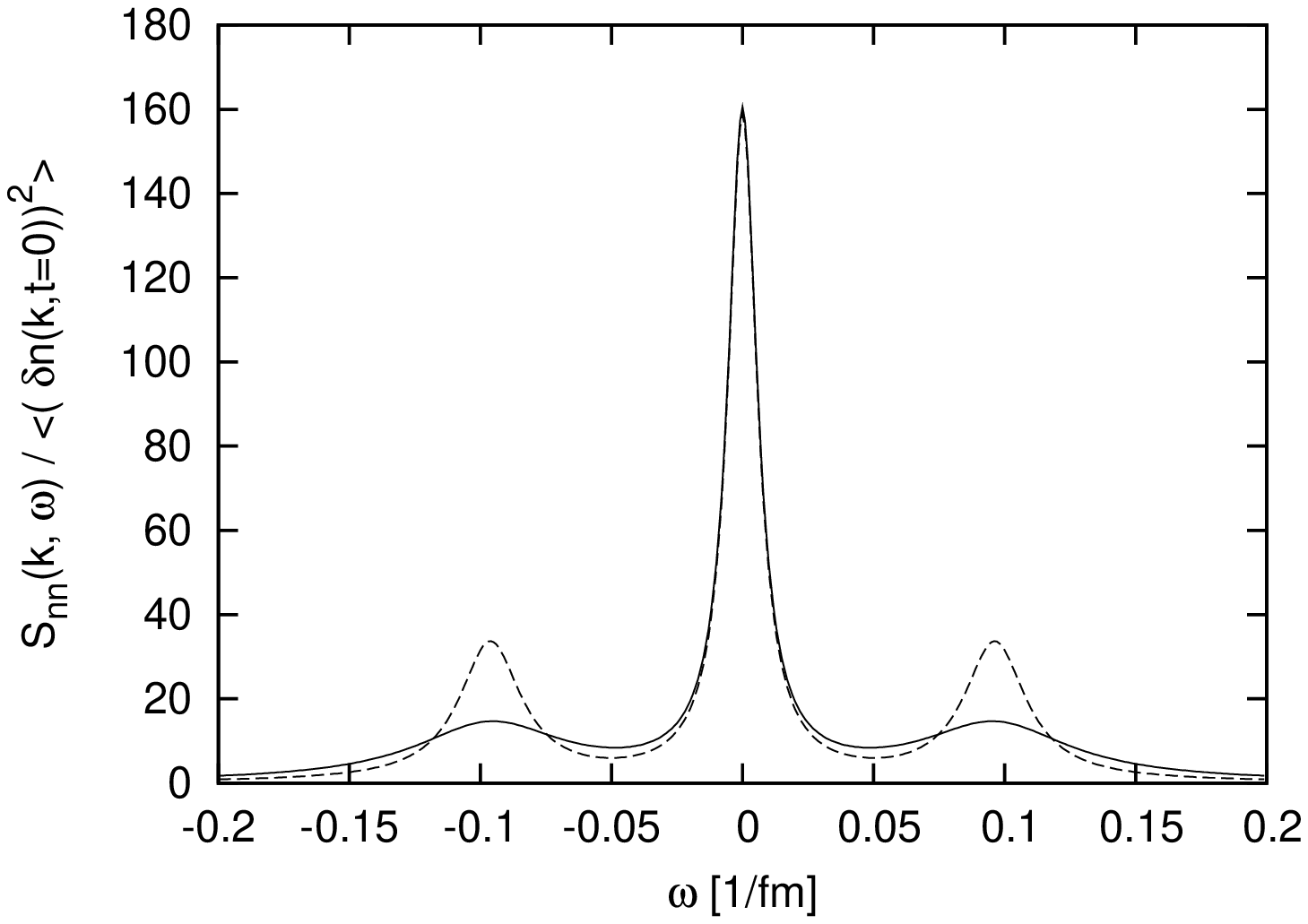}
\end{center}
\end{minipage}
\begin{minipage}{0.5\hsize}
 \begin{center}
   \includegraphics[width=40mm,angle=-90]{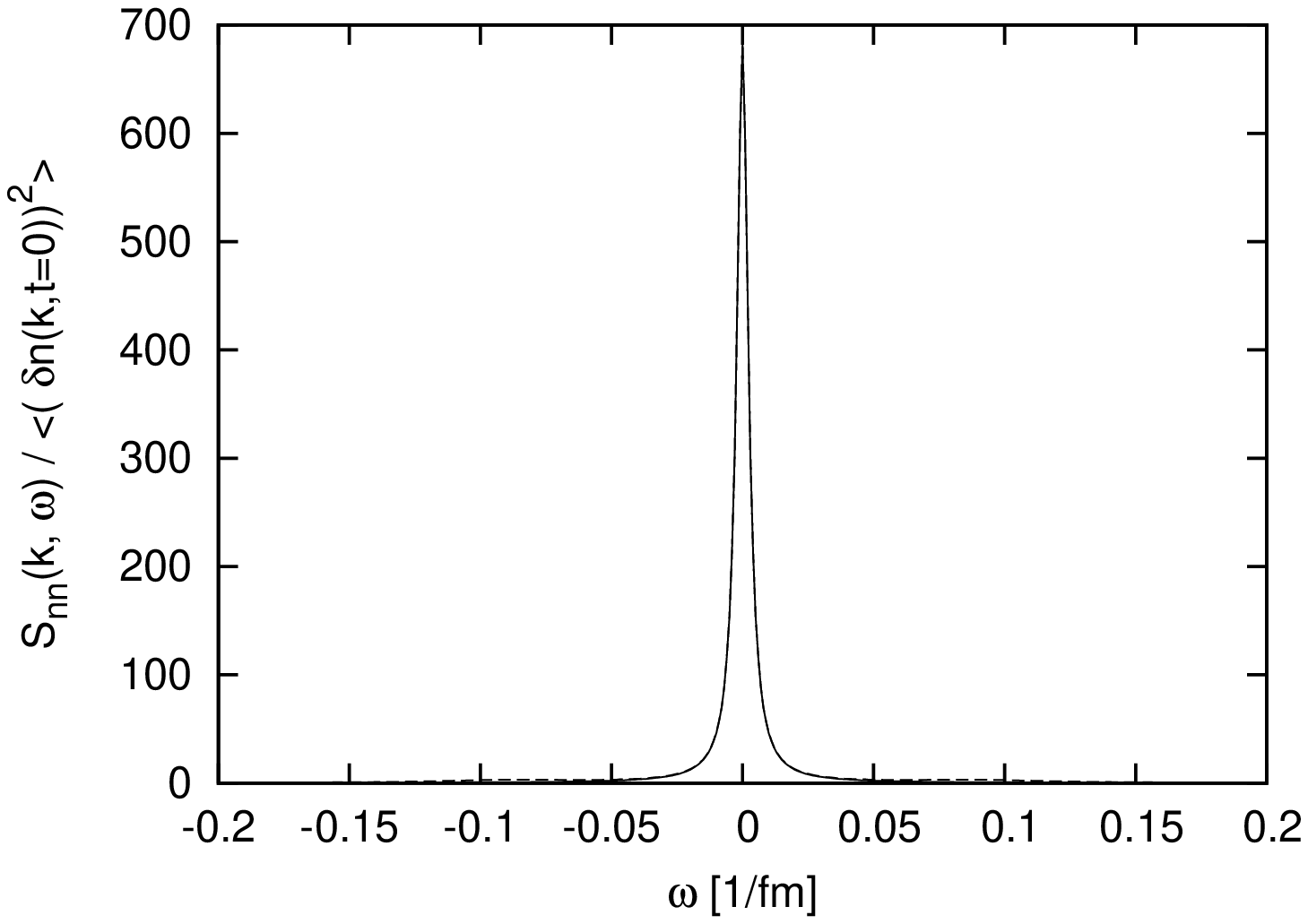}
\end{center}
\end{minipage}
\end{tabular}
  \caption{{\bf Left panel:}\,
The spectral function at $t\equiv (T-T_c)/T_C=0.5$ and $k=0.1$\, [1/fm].
               The solid line represents the results using
the  Landau equation (energy frame) and Israel-Stewart equation.
           The dashed line represents the result using a fluid dynamic equation
in the particle frame\cite{tko}. 
           The strength of the Brillouin peaks becomes small due to the singularity of
 the ratio of specific heats.
 {\bf Right panel:}\, The spectral function at $t=0.1$ and $k=0.1$\, [1/fm].
 The sound modes die out when approaching the CP,  irrespective of the
relativistic fluid dynamic equations used. Taken from \citen{Minami:2009hn}.
 }
  \label{fig:t5-1}
%
\end{figure}

Such an attenuation of the mechanical density fluctuations
 may lead to a suppression or even total disappearance
 of  Mach cone at the QCD CP\cite{Minami:2009hn}.
 If the Mach cone formation is confirmed at some incident
energy in relativistic heavy-ion collisions,
 possible disappearance or strong suppression of a Mach cone along 
with the lowering
of the incident energy can be a signal of the existence of the critical point,
because it may mean that 
the created matter should have gone through the critical region of the CP.

Eexplicit calculations with equation of motion which admits the existence
of the critical point is necessary for confirm the fate of Mach cone formation. 
To make a direct connection with experimental observations, we should analyze the
density fluctuations with the expanding back ground.

\section{Summary and concluding remarks}

In the first part,
after emphasizing  that the  repulsive vector interaction should exist
between the quarkswe, we have shown that there is 
a possibility that  the QCD phase diagram may have  multiple
 critical points when the color superconductivity 
and the vector interaction are incorporated in the mean-field level.  
The message of this findings is
 that the QCD matter in the vicinity of the phase boundary at low or moderate
temperature is very soft for the formation
of  diquark and chiral condensates combined with the baryonic density,
as described by $a\bar{q}^cq+b\bar{q}q+ cq^{\dag}q$, along the critical line. 
 An analysis of such a possibility
may involve a {\em dynamical} Hartre(-Fock)-Bogoliubov theory in the
relativistic kinematics, which should be an interesting theoretical challenge.

It was also shown that the vector interaction as given by
Eq.(\ref{Lagrangian2f}) 
 suppresses the chromomagnetic instability related to
asymmetric homogeneous 2CSC phase:
 With increasing vector interaction,
the unstable region associated with chromomagnetic instability
shrinks towards lower temperature and higher chemical potential.
That means that the vector interaction can at least partially 
resolve  the chromomagnetic instability problem.

The dynamical density fluctuations have been analyzed using relativistic fluid dynamics,
in which the entropy fluctuations are automatically incorporated.
The sound modes due to density fluctuations are attenuated,
 and the Rayleigh peak due to the entropy fluctuation in turn gets enhanced 
around the QCD critical point.
 The attenuation of the sound mode may lead to the suppression or even total disappearance 
of  Mach cone at the CP.
 Once the Mach cone formation is confirmed in experiments of heavy-ion collisions
with a high incident energy,
possible disappearance or strong suppression along with the variation of 
the initial energy can be a signal of the existence of the critical point 
and the created matter went through the critical region of the CP.
 It is clear that there need further explicit calculations for a 
confirmation of this conjecture.

The present analysis is made in the fluid dynamical regime.
In the very vicinity of the critical point,
we need an analysis beyond the fluid dynamics to
take into account  the non-linear effects. For this purpose,
the  mode-mode coupling theory\cite{Kawasaki:1976zz}
and/or dynamical renormalization group technique \cite{Hohenberg:1977ym}
 should be applied\cite{onuki,progress}. 

There are other  QCD phase transitions than those discussed 
in the present report, and if the phase transition  is of a second order or close to that, 
there should exist specific soft modes, which may be easily thermally excited.
Such a bosonic excitation with a small mass could
in turn affect the quark spectral functions:
In the high-temperature phase above the critical 
point $T_c$ of the color super conductivity,
the quark spectral function naturally exhibits a pseudo gap\cite{Kitazawa:2003cs}
 owing to the
coupling with the pre-formed diquark fluctuations above $T_c$\cite{Kitazawa:2001ft}.
The chiral soft modes\cite{Hatsuda:1984jm} may also lead to a complicated quark 
spectral function with three-peak structure above but around the
critical temperature of the chiral transition\cite{Kitazawa:2005mp}.

\section*{Acknowledgements}
This work was partially supported by a
Grant-in-Aid for Scientific Research by the Ministry of Education,
Culture, Sports, Science and Technology (MEXT) of Japan (No.
20540265),
 by Yukawa International Program for Quark-Hadron Sciences, and by the
Grant-in-Aid for the global COE program `` The Next Generation of
Physics, Spun from Universality and Emergence '' from MEXT.

%

\end{document}